\RequirePackage{fix-cm}
\documentclass[twocolumn,epjc3]{svjour3}
\smartqed  
\RequirePackage{graphicx}
\usepackage{amssymb}
\journalname{Eur. Phys. J. C}
\begin{document}

\title{Testing generalized spacetimes as black holes using Hod function for the hoop conjecture}


\author{K.K. Nandi\thanksref{e1,addr1,addr2}
        \and
        R.N. Izmailov\thanksref{e2,addr1}
        \and
        R.Kh. Karimov\thanksref{e3,addr1}
        \and
        G.M. Garipova\thanksref{e4,addr1}
        \and
        R.R. Volotskova\thanksref{e5,addr1,addr3}
        \and
        A.A. Potapov\thanksref{e6,addr2}
}

\thankstext{e1}{e-mail: kamalnandi1952@rediffmail.com}
\thankstext{e2}{e-mail: izmailov.ramil@gmail.com}
\thankstext{e3}{e-mail: karimov\_ramis\_92@mail.ru}
\thankstext{e4}{e-mail: goldberg144@mail.ru}
\thankstext{e5}{e-mail: fmfizika15@mail.ru}
\thankstext{e6}{e-mail: a.a.potapov@strbsu.ru}


\institute{Zel'dovich International Center for Astrophysics, M. Akmullah Bashkir State Pedagogical University, 3A,
           October Revolution Street, Ufa 450008, RB, Russia \label{addr1}
           \and
           Department of Physics \& Astronomy, Bashkir State University, 47A, Lenin Street, Sterlitamak 453103, RB, Russia \label{addr2}
           \and
           Salavat Industrial College, 27, Matrosova Boulevard, Salavat 453259, RB, Russia \label{addr3}
}

\date{Received: date / Accepted: date}

\maketitle

\begin{abstract}
The hoop conjecture, due to Thorne, is a fundamental aspect of black holes in classical general relativity. Recently, generalized classes of regular spherically symmetric static black holes with arbitrary exponents coupled to nonlinear electrodynamics have been constructed in the literature. The conjecture in those spacetimes could be violated if only the asymptotic mass $M_{\infty}$ is used. To avoid such violations, Hod earlier suggested the appropriate mass function and stated the conjecture in terms of what we call the Hod function. The conjecture can then be applied to any given static spacetime to test whether or not it represents black holes. It is shown here that the conjecture is protected in the above constructed class of generalized spacetimes thus supporting them as black holes. However, it is argued that there are factors, including violation of the conjecture, that militate against the proposed \textit{new} class of solutions to be qualifying as black holes. Finally, we exemplify that the Hod mass $M(r\leq R)$ in the conjecture is exactly the \textit{matter} counterpart of the Misner-Sharp \textit{geometrical} quasilocal mass $m(r\leq R)$ of general relativity. Thus any conclusion based on Hod function is strictly a conclusion of general relativity.
\end{abstract}

\section{Introduction}
\label{intro}
The hoop conjecture \cite{Thorne:1972,Misner:1973} asserts that a self-gravitating matter configuration of mass $M$ will form an engulfing horizon if its circumferential radius $R=\frac{C}{2\pi }$ is equal to (or less than) the corresponding Schwarzschild radius $2M$ (in units $G=1,c=1$) from all directions. That is, the conjecture states that a black hole horizon exists if
\begin{equation}
C\leq 4\pi M\Rightarrow \textmd{black hole horizon}.
\end{equation}

Several studies have supported this relation (see, e.g., \cite{Redmount:1983,Bizon:1988,Abrahams:1992,Zannias:1993,Chiba:1994,Chiba:1999,Nakao:2003,Khuri:2009,Cvetic:2011,Malec:2015,Hod:2015,Ma:2015,Barrow:2017,Anza:2017,Peng:2019,Hod:2019}). Nevertheless, there has been some claims in the literature that the hoop conjecture can be violated in the linearly coupled charged spacetimes \cite{deLeon:1987,Bonnor:1983} but it is naively based on using the definition of an asymptotic mass. To avoid such violations, Hod \cite{Hod:2018} suggested that \textit{the mass term on the r.h.s of (1) should be interpreted as the gravitational mass} $M(r\leq R)$ \textit{contained within the engulfing hoop of radius} $R$ \textit{and not as the total (asymptotically measured) mass} $M_{\infty}$ \textit{of the entire spacetime.} The mass $M(r\leq R)$ summarizes the conjecture in terms of what we call Hod function $H$, which was first used in \cite{Nandi:2020}.

Fan and Wang \cite{Fan:2018} constructed generalized arbitrary two-parameter family of spherically symmetric asymptotically flat black hole solutions of general relativity coupled to nonlinear electrodynamics that are characterized by a Schwarzschild mass $M_{\scriptsize{\textmd{Sch}}}$, asymptotic electromagnetic mass $M_{e}$, magnetic charge $Q_{m}$ or electric charge $Q_{e}$, and two arbitrary constant exponents $\mu$ and $\nu$. When $M_{\scriptsize{\textmd{Sch}}}=0$, these solutions represent black hole spacetimes that are built purely by the nonlinear electromagnetic field and are regular everywhere including at the origin (centrally regular). When $M_{\scriptsize{\textmd{Sch}}}\neq 0$, the spacetime has a singularity at the origin but regular elsewhere\footnote{%
Originally, what Fan and Wang \cite{Fan:2018} denoted by $M$ is denoted here by $M_{\scriptsize{\textmd{Sch}}}$ in order to distinguish it from the ADM mass $M_{\infty }$, which is given by $M_{\scriptsize{\textmd{Sch}}}+M_{e}$ [see Eq.(6) below]. When $M_{e}=0$, the metric becomes Schwarzschild [see metric (4)], hence the nomenclature $M_{\scriptsize{\textmd{Sch}}}$ but it is essentially the same as $M$ of Fan and Wang. The presence of $M_{\scriptsize{\textmd{Sch}}}\neq 0$ in metric (4) introduces a singularity at $r=0$. On the other hand, the solution with $M_{\scriptsize{\textmd{Sch}}}=0$ but $M_{e}\neq 0$ is an everywhere (including at $r=0$) regular solution.}. Thus the solutions are interesting by themselves deserving further studies.

In this paper, we study if the Hod function, extended to cover complicated generalized solutions of general relativity, can test whether or not those solutions truly represent black holes. It turns out that the spacetimes constructed by Fan and Wang respect the hoop conjecture so that they genuinely represent black holes. However, their proposed \textit{new} class of solutions \cite{Fan:2018} is shown to not respect the conjecture. We shall also exemplify, using those solutions, that Hod's definition of $M(r\leq R)$ is exactly the matter counterpart of the Misner-Sharp \cite{Redmount:1983,Misner:1964} geometrical quasilocal mass $m(r\leq R)$ of general relativity.

The paper is organized as follows: In Sec.2, we define the critical values of the physical parameters separating horizon and no horizon regimes. In Sections 3 and 4, we develop the extended Hod function for the Fan and Wang solutions for $M_{\scriptsize{\textmd{Sch}}}=0$ and $M_{\scriptsize{\textmd{Sch}}}\neq 0$ respectively and restate the hoop conjecture as constraints on the extended Hod function. These constraints are examined for the considered two parameter solutions. In Sec.5, the proposed new solution is studied including its curvature properties. In Sec.6, we exemplify that the equality of Hod mass and the Misner-Sharp mass. Sec.7 summarizes the paper.

\section{Fan and Wang generic class of solutions}
\label{sec:2}
Fan and Wang \cite{Fan:2018} spacetimes under consideration follow from the gravitational action $S$ with source of nonlinear electrodynamics%
\begin{equation}
S=\int \sqrt{-g}d^{4}x\left[ \frac{1}{16\pi }R-\frac{1}{4\pi }\mathcal{L}(F)\right],
\end{equation}%
where $R$ is the Ricci scalar, $F = \frac{1}{4} F_{\mu\nu} F^{\mu\nu}$ and $\mathcal{L}(F)$ is a nonlinear electrodynamic Lagrangian\footnote{%
We should state that the $\mathcal{L}(F)$ formalism has "underwater stones" \cite{Bronnikov:2000}, hence may not be appropriate for these spacetimes since the Lagrangian is multivalued having different branches. There is however an alternative elegant Hamiltonian formalism with $\mathcal{H}(P)$ obtained by the Legendre transformation \cite{Plebanski:1968}, where $P = F \left(\frac{\partial \mathcal{L}}{\partial F}\right)^{2}$. It is important to emphasize, following Novello et al \cite{Novello:2000} that the ambiguity in the Lagrangian is reflected in the appearance of singularities in the "effective" spacetime built by electromagnetic nonlinearities. However, these singularities are felt \textit{only} by photons, while the rest of the matter follows usual smooth geodesics. Since the hoop conjecture is related to black holes formed by geodesically collapsing matter (and not by photons), such singularities do not pose any problem as long as the spacetime stresses satisfy the needed energy conditions required for collapse.}.

Consider their generic metric \cite{Fan:2018}
\begin{eqnarray}
d\tau^{2} &=& -f(r)dt^{2}+\frac{1}{f(r)}dr^{2}+r^{2}(d\theta ^{2}+\sin^{2}\theta d\psi ^{2}), \\
f(r) &=& 1-\frac{2M_{\scriptsize{\textmd{Sch}}}}{r}-\frac{2M_{e}r^{\mu -1}}{\left( r^{\nu}+q^{\nu}\right)^{\mu /\nu}},
\end{eqnarray}%
where $\mu,\nu$ are arbitrary constant exponents, $q>0$ is related to the magnetic or electric charge and
\begin{equation}
M_{e}=\frac{q^{3}}{\alpha}
\end{equation}%
is the asymptotic electromagnetic ADM mass, $\alpha >0$ has the dimension of (length)$^{2}$. The asymptotic total ADM mass $M_{\infty}$ is%
\begin{equation}
M_{\infty }=M_{\scriptsize{\textmd{Sch}}}+M_{e}.
\end{equation}%
As such, analytically finding horizon radii for higher arbitrary values of $\mu ,\nu$ is difficult, if not impossible. They can of course be found numerically, but we need more information beyond these radii to study the hoop conjecture.

To obtain them, we define, following the notations of Ay\'{o}n-Beato and Garc\'{\i}a \cite{Ayon-Beato:1998}, the following rescaled quantities
\begin{equation}
x=\frac{r}{q},s=\frac{q}{2M_{e}},w=\frac{M_{\scriptsize{\textmd{Sch}}}}{q},
\end{equation}%
then
\begin{equation}
f(x,s,w)=1-\frac{2w}{x}-\frac{x^{\mu -1}\left( 1+x^{\nu }\right) ^{-\mu /\nu
}}{s},
\end{equation}%
where $x>0$ is the only variable, the rest are constant parameters. Note that $q$ cannot be negative since the "standard" coordinate radius $r\in \lbrack 0,\infty )$. The critical values, or minima, $x_{c}$ and $s_{c}$, are obtained by solving the simultaneous equations
\begin{eqnarray}
f(x_{c},s_{c},w) &=&0, \\
\partial _{x}f(x_{c},s_{c},w) &=&0.
\end{eqnarray}%
These equations suggest that the existence of horizon provides an upper limit $w_{c}$ beyond which $x_{c},s_{c}$ cannot be defined.

\begin{figure}[!ht]
  \centerline{\includegraphics[scale=1.05]{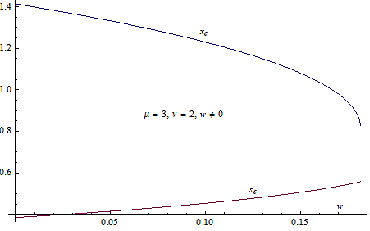}}
  \caption{Plot for $x_{c}(w)$ and $s_{c}(w)$ vs $w$ for $\mu =3,\nu =2$.}
  \label{fig1}
\end{figure}

For illustration, we choose the Bardeen solution \cite{Bardeen:1968} with values $\mu =3$, $\nu =2$ with $w\neq 0$. (See also \cite{Kodama:1980}). This case is also helpful in the sense that the results of the original Bardeen solution $w=0$ (or $M_{\scriptsize{\textmd{Sch}}}=0$) can be retrieved. The Eqs.(9) and (10) yield a single fixed pair of values $\left( x_{c},s_{c}\right) $:%
\begin{eqnarray}
x_{c} &=&\frac{2\times 3^{1/3}+\left( -27w+\sqrt{-24+729w^{2}}\right) ^{2/3}}{3^{2/3}\left( -27w+\sqrt{-24+729w^{2}}\right) ^{1/3}} \\
s_{c} &=&3x_{c}^{2}\left( 1+x_{c}^{2}\right) ^{-5/2}.
\end{eqnarray}%
Thus, $x_{c},s_{c}$ are defined only for $w\leq w_{c}=\sqrt{24/729}=0.1814$. It follows from Fig.1 that the interval $w\in \lbrack 0,\sqrt{24/729}]$ corresponds to $x_{c}\in \lbrack \sqrt{2},0.8164]$ and $s_{c}\in \lbrack \frac{2}{3\sqrt{3}},0.5577]$. For any choice of $w\leq w_{c}$, we obtain the corresponding critical values $x_{c},s_{c}$ from Eqs.(11,12). Then, by varying $x$, we can plot $f(x,s,w)$, where the parameter $s$ is allowed to assume values in positive multiples of $s_{c}$ for any choice of fixed legal $w$. We choose for illustration $w=0$ and $0.05$ in Secs.3 and 4, respecting the constraint $w\leq w_{c}$.

The transition between the existence of "no horizon and horizon" is geometrically marked by the critical parameter $s_{c}$ for $f(x,s,w)$ as follows \cite{Ayon-Beato:1998}:%
\begin{eqnarray}
s &>&s_{c}\textmd{ }\Rightarrow \textmd{ no horizon} \\
s &<&s_{c}\textmd{ }\Rightarrow \textmd{ black hole horizon} \\
s &=&s_{c}\textmd{ }\Rightarrow \textmd{ two coincident horizons.}
\end{eqnarray}%
This transition is illustrated in Figs.2,3 ($w=0$) and in Figs.4,5 ($w=0.05$) with values of $\mu ,\nu $ specified there. The plots of $f(x,s,0.05)$ show similar patterns for different values of $w,\mu ,\nu $ as well, so not shown in this paper.

Let us turn to the hoop conjecture for $w=0$. It could be violated if one na\"{\i}vely uses just the asymptotic mass $M_{\infty }=M_{e}$ and write the conjecture as%
\begin{equation}
\frac{C(R)}{4\pi M_{e}}=\frac{R}{2M_{e}}=sx.
\end{equation}%
Note that the existence of horizon has been described in \cite{Ayon-Beato:1998} by the curves for $f(x,s,w)$\ plotted for $x\in \lbrack 0,\infty ]$, each curve being parametrized by $s$ that takes on values in positive multiples of $s_{c}$. Consider any curve, say, the Hayward solution \cite{Hayward:2006} ($\mu =3,\nu =3)$ in Fig.3, where $x_{c}=1.2599$, $s_{c}=0.5291$ and take a curve in the "no-horizon" regime designated by $s=0.6$ (say). Then one obtains the value on the right hand side of Eq.(16) as $sx_{c}=0.7559<1$ suggesting that there would be a horizon, while there is none. This seemingly amounts to a \textit{violation} of the conjecture, but that would be a misconception. Using the generic Hod function, derived below, we can restore the validity of the conjecture respecting the inequalities (13-15).

We emphasize that the entire analysis can be performed for \textit{any} value of $\mu, \nu$ and for a legal choice of $w\leq w_{c}$, where $w_{c}$ is determined by the requirement that $x_{c},s_{c}$ be real. This constraint $M_{\scriptsize{\textmd{Sch}}}\leq qw_{c}$ is similar to what one finds in the Reissner-Nordstr\"{o}m black hole, suggesting that $M_{\scriptsize{\textmd{Sch}}}$ can no longer be freely specified if we want the metric to have horizons. There are no black holes for $M_{\scriptsize{\textmd{Sch}}}>$ $qw_{c}$.

\section{Hod function ($M_{\scriptsize{\textmd{Sch}}}=0, w=0$)}
\label{sec:3}

The energy outside a charged ball of radius $R$ is
\begin{equation}
E_{\scriptsize{\textmd{elec}}}(r>R)=\frac{Q^{2}}{2R},
\end{equation}%
where $\rho (r>R)=\frac{Q^{2}}{8\pi r^{4}}$ is the electric energy density in linear electrodynamics. According to the interpretation by Hod \cite{Hod:2018}, stated already in Sec.1, the gravitational mass $M(r\leq R)$ contained within ($r\leq R$) of the ball is given by%
\begin{equation}
M(r\leq R)=M_{\infty }-\frac{Q^{2}}{2R},
\end{equation}%
and this mass (we call it Hod mass) should be used instead of the asymptotic ADM mass $M_{\infty}$ in (1). With this, it was shown in \cite{Hod:2018} that the example of a curved spacetime coupled to linear electrodynamics, though seemingly violate the hoop conjecture, actually obeys it.

Since we are concerned in this paper with generic Fan and Wang \cite{Fan:2018} spacetimes coupled to nonlinear electrodynamics with arbitrary exponents $\mu ,\nu$, we shall extend the formula for the Hod mass (18) integrating the corresponding $T_{0}^{0}$ of a given solution, that is,
\begin{equation}
M(r\leq R)=M_{\infty }-E_{\scriptsize{\textmd{elec}}}=M_{\infty }-\frac{1}{8\pi }%
\int_{R}^{\infty }T_{0}^{0}4\pi r^{2}dr.
\end{equation}%
When $M_{\scriptsize{\textmd{Sch}}}=0$ or $w=0$, we have the asymptotic ADM mass for the entire space to be $M_{\infty }=M_{e}$. The energy density $T_{0}^{0}$ for the metric (8) can be obtained from the field equations $\left(\frac{1}{8\pi}\right) G_{0}^{0}g^{00}=T_{0}^{0}$, where $G_{0}^{0}$ is the Einstein tensor. Working out $T_{0}^{0}$ and integrating it, we find%
\begin{equation}
E_{\scriptsize{\textmd{elec}}}=\frac{1}{8\pi }\int_{R}^{\infty }T_{0}^{0}4\pi r^{2}dr = M_{e}\left[ 1-\left\{ 1+\left( \frac{q}{R}\right) ^{\nu }\right\} ^{-\mu /\nu }\right],
\end{equation}%
Then, with $\frac{M_{e}}{R}=\frac{1}{2sx}$, $x=\frac{R}{q}$, we obtain
\begin{eqnarray}
\frac{M(r\leq R)}{R} &=&\frac{M_{e}-E_{\scriptsize{\textmd{elec}}}}{R} \\
&=&\frac{\left[ 1+\left( \frac{1}{x}\right) ^{\nu }\right] ^{-\mu /\nu }}{2sx}.
\end{eqnarray}%
With this, we introduce what we call the Hod function $H_{f}(\mu ,\nu ;x,s)$ for the metric (8):%
\begin{eqnarray}
\frac{C(R)}{4\pi M(r\leq R)} &=&sx\left[ 1+\left( \frac{1}{x}\right) ^{\nu }\right] ^{\mu /\nu } \\
&\equiv &H_{f}(\mu ,\nu ;x,s).
\end{eqnarray}%
Note that $H_{f}(\mu ,\nu ;x,s)>0$ since the total asymptotic electromagnetic mass for the whole space is $M_{e}$ so that it is bigger than the electromagnetic energy $E_{\scriptsize{\textmd{elec}}}$ enclosed within part of the space. The Eqs.(9,10) yield for the metric (8)%
\begin{equation}
x_{c}=(\mu -1)^{1/\nu }\textmd{, }s_{c}=\mu ^{-\mu /\nu }\left[ (\mu
-1)^{1/\nu }\right] ^{\mu -1}.
\end{equation}%
A typical plot of the\ Bardeen solution \cite{Bardeen:1968} $f(\mu =3,\nu =2;x,s_{c})$ is displayed in Fig.2. The uppermost curve represents "no horizon" regime for $s>s_{c}$, middle curve represents coincident or extreme horizon at the transitional value $s=s_{c}$ and the lowest curve represents horizon that occurs for $s<s_{c}$. The Hayward solution \cite{Hayward:2006} $f(\mu =3,\nu =3;x,s)$ is displayed in Fig.3. Thus, the Bardeen and Hayward solutions are confirmed as true black holes. We expect that the existence of horizon would be marked by the Hod function (24) for arbitrary exponents $\mu ,\nu $ as follows:
\begin{eqnarray}
0 &<&H_{f}(\mu ,\nu ;x,s)\leq 1\textmd{ for }s\leq s_{c}\textmd{ }\Rightarrow
\textmd{ horizon} \\
&&H_{f}(\mu ,\nu ;x,s)  > 1\textmd{ for }s>s_{c}\textmd{ }\Rightarrow \textmd{ no
horizon.}
\end{eqnarray}%
Fig.6 illustrates that this is indeed the case. \textit{The above inequalities are nothing but restatements of Thorne's hoop conjecture defined by the Hod mass (19), which is identical to the Misner-Sharp quasilocal mass of general relativity, as will be exemplified later.} Therefore, these inequalities must be satisfied by any solution for it to qualify as a black hole.

\begin{figure}[!ht]
  \centerline{\includegraphics[scale=1.05]{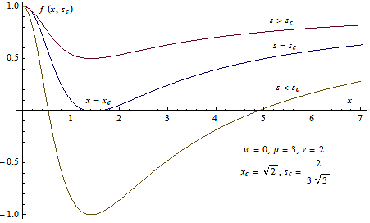}}
  \caption{Behavior of $f(x;s)$ for fixed values of parameter $s<s_{c}$ and $w=0$, $\mu =3,\nu =2$.}
  \label{fig2}
\end{figure}

\begin{figure}[!ht]
  \centerline{\includegraphics[scale=1.05]{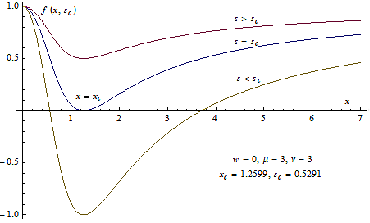}}
  \caption{Behavior of $f(x;s)$ for fixed values of parameter $s<s_{c}$ and $w=0$, $\mu =3,\nu =3$.}
  \label{fig3}
\end{figure}

\begin{figure}[!ht]
  \centerline{\includegraphics[scale=1.05]{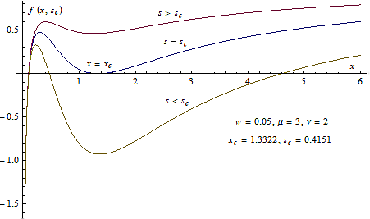}}
  \caption{Behavior of $f(x;s)$ for fixed values of parameter $s<s_{c}$ and $w=0.05$, $\mu =3,\nu =2$.}
  \label{fig4}
\end{figure}

\begin{figure}[!ht]
  \centerline{\includegraphics[scale=1.05]{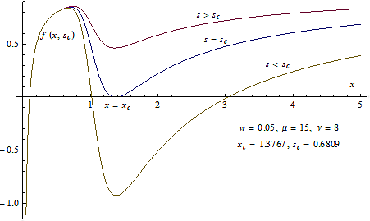}}
  \caption{Behavior of $f(x;s)$ for fixed values of parameter $s<s_{c}$ and $w=0.05$, $\mu =15,\nu =8$.}
  \label{fig5}
\end{figure}

\begin{figure}[!ht]
  \centerline{\includegraphics[scale=1.05]{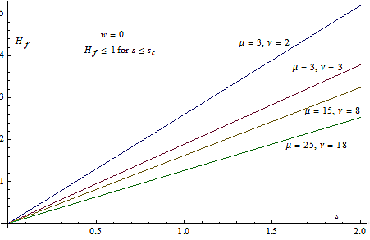}}
  \caption{Behavior of the extended Hod function $H_{f}(x_{c},s)$ for $w=0$ for various values of $\mu $ and $\nu $. We display the critical values for the reader to verify that the function truly embodies the hoop conjecture. The values are: $x_{c}=1.414,s_{c}=0.3839$ ($\mu =3,\nu =2$); $x_{c}=1.2599,s_{c}=0.5291$ ($\mu =3,\nu =3$); $x_{c}=1.3908,s_{c}=0.6317$ ($\mu =15,\nu =8$); $x_{c}=1.931,s_{c}=0.7919$ ($\mu =25,\nu =18$).}
  \label{fig6}
\end{figure}

\section{Hod function ($M_{\scriptsize{\textmd{Sch}}}\neq 0, w=0.05$)}
\label{sec:4}

We next use the metric (8) with $M_{\scriptsize{\textmd{Sch}}}\neq 0$ or $w\neq 0$. It was shown in Sec.2 that $w$ has always an upper limit $w_{c}$ coming from the constraint of reality of $x_{c}$ and $s_{c}.$ For arbitrary choices of $\mu,\nu$, the expressions for $x_{c}$ and $s_{c}$ become messy but it is possible to choose some small consistent value of $w$ protecting the metric signature without running into imaginary quantities in the subsequent calculations. We choose $w=0.05$ for illustration only but is otherwise subject only to $w\leq w_{c}$.

Let us turn to the hoop conjecture for $w\neq 0$. It could be violated if one uses just the asymptotic mass $M_{\infty }=M_{\scriptsize{\textmd{Sch}}}+M_{e}$ leading to the conjecture as%
\begin{equation}
\frac{C(R)}{4\pi M_{\infty }}=\frac{sx}{1+2sw}.
\end{equation}%
Consider the Bardeen solution $\mu =3,\nu =2$, with $w=0.05$ (Fig.4), where $x_{c}=1.3322$, $s_{c}=0.4151$ and take a curve in the "no-horizon" regime designated by $s=0.5$ (say). Then one obtains, for fixed $x_{c}$,\ the value on the right hand side of Eq.(28) as $\frac{C(R)}{4\pi M_{\infty }}=0.6343<1$ suggesting that there would be a horizon, while there is none. This amounts to a violation of the conjecture. Again this violation is ruled out by the Hod function derived below.

The energy density $T_{0}^{0}$ for the metric (8) can be obtained from the field equations but surprisingly we find that it is \textit{independent} of $M_{\scriptsize{\textmd{Sch}}}$ and its integration leads to the same $E_{\scriptsize{\textmd{elec}}}$ as given by Eq.(20). Proceeding as before, we obtain%
\begin{eqnarray}
\frac{M(r\leq R)}{R} &=&\frac{M_{\infty }-E_{\scriptsize{\textmd{elec}}}}{R} \\
&=&\left( \frac{1}{x}\right) \left[ w+\frac{1}{2s}\left\{ 1+\left( \frac{1}{x%
}\right) ^{\nu }\right\} ^{-\mu /\nu }\right] ,
\end{eqnarray}%
which leads to the more general Hod function with a new parameter $w$ as follows%
\begin{equation}
H_{f}(w,\mu ,\nu ;x,s)=\frac{sx}{2sw+\left\{ 1+\left( \frac{1}{x}\right)
^{\nu }\right\} ^{-\mu /\nu }}.
\end{equation}%
We expect $H_{f}(w,\mu ,\nu ;x,s)$ to satisfy the inequalities, viz.,%
\begin{eqnarray}
0 &<&H_{f}(w,\mu ,\nu ;x,s)\leq 1\textmd{ for }s\leq s_{c}\textmd{ }\Rightarrow
\textmd{ horizon} \\
&>&1\textmd{ for }s>s_{c}\textmd{ }\Rightarrow \textmd{ no horizon.}
\end{eqnarray}%
if Thorne's conjecture is to be protected so that the solution has horizons. The functions $H_{f}(0.05,\mu ,\nu ,x_{c};s)$ are shown in Fig.7 for some illustrative values of $\mu ,\nu $. We have taken $x=x_{c}$ which show that the inequalities are exactly satisfied for the corresponding $s_{c}$ being the transition point, meaning that the conjecture is protected for the coincident horizon. This conclusion holds for any value of $\mu ,\nu $, and $w$ that respects the reality of $s_{c}$ and $x_{c}$ given by Eqs.(11) and (12).

\begin{figure}[!ht]
  \centerline{\includegraphics[scale=1.05]{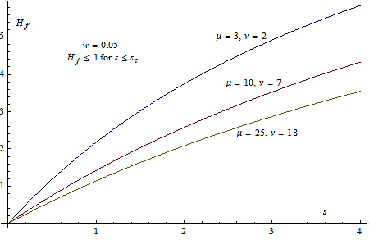}}
  \caption{Behavior of the extended Hod function $H_{f}(x_{c},s)$ for $w=0.05$ for various values of $\mu $ and $\nu $. We display the critical values for the reader to verify that the Hod function truly embodies the hoop conjecture. The values are: $x_{c}=1.3322,s_{c}=0.4151$ ($\mu =3,\nu =2$); $x_{c}=1.352,s_{c}=0.6873$ ($\mu =10,\nu =7$); $x_{c}=1.1870,s_{c}=0.8646$ ($\mu =25,\nu =18$).}
  \label{fig7}
\end{figure}

\section{The new class of Fan-Wang solutions}
\label{sec:5}

This new class of black hole solutions has the virtue that the vector field approaches a Maxwell field in the weak field limit, which is certainly a desirable aspect. However, the solutions possess some peculiarities as detailed below. The metric is ($\alpha >0$, $q>0$, as assumed in \cite{Fan:2018})

\begin{eqnarray}
d\tau^{2} &=& -g(r)dt^{2} + \frac{1}{g(r)}dr^{2} + r^{2}(d\theta^{2} + \sin^{2}\theta d\psi ^{2}), \\
g(r) &=& 1-\frac{2M_{\scriptsize{\textmd{Sch}}}}{r} - \frac{2M_{e}r^{\mu -1}}{\left(r+q\right)^{\mu }}..
\end{eqnarray}%
Using the rescaling as in (7), we rewrite the metric as%
\begin{equation}
g(x;s,w)=1-\frac{2w}{x}-\frac{\left( 1+x\right) ^{-\mu }}{xs}.
\end{equation}%
For the metric form (35), we find%
\begin{equation}
E_{\scriptsize{\textmd{elec}}}=\frac{1}{8\pi }\int_{R}^{\infty }T_{0}^{0}4\pi r^{2}dr=M_{e}%
\left[ 1-\frac{R^{\mu }}{(q+R)^{\mu }}\right]
\end{equation}%
and the corresponding Hod function $H_{g}$ becomes%
\begin{equation}
H_{g}\left( w,\mu ,x,s\right) =\frac{xs}{2ws+x^{\mu }(1+x)^{-\mu }}.
\end{equation}%
For simplicity, let us consider the case $\mu =3$. The conditions (9,10)
then yield%
\begin{equation}
x_{c}=\frac{1}{4}\left( -1+6w\right) \textmd{, }s_{c}=-\frac{256}{27(1+2w)^{4}}%
.
\end{equation}%
It is evident that whatever be the value of $w$, the value of $s_{c}$ is always negative, which is inconsistent with the stipulated definition
\begin{equation}
s=\frac{q}{2M_{e}}=\frac{\alpha }{2q^{2}}>0.
\end{equation}

Consider the value $w=0$ or $M_{\scriptsize{\textmd{Sch}}}=0$. This case yields
\begin{equation}
x_{c}=-\frac{1}{4},s_{c}=-\frac{256}{27}.
\end{equation}%
The negative value of $x_{c}$ is unphysical since radius $r$ ($\equiv xq$) in "standard coordinates" cannot be negative by definition, which rules out the case $w=0$ as black holes. For curiosity, we allow negative values of $x$, then Fig.8 shows the complete picture for all $-\infty <x<\infty $ with the parameter $s$ taking on values in positive multiples of $s_{c}$ in the metric (36). The pattern of the plot of (36) is the same whatever be the value $\mu $. Clearly, there are no horizons within the valid coordinate range $0<x<\infty $. In any case, $x_{c}>0$ requires that $w>\frac{1}{6}$, which already imposes a constraint $6M_{\scriptsize{\textmd{Sch}}}>q$ for which there are no obvious physical grounds.

Consider now the value $w=1$ or $M_{\scriptsize{\textmd{Sch}}}\neq 0$. This case yields positive $x_{c}=\frac{5}{4}$ but $s_{c}$ is still negative ($s_{c}=-\frac{256}{2187}$). The signs of $x_{c}$ and $s_{c}$ remain the same no matter what the value of $w$ or $\mu $ is, as can be easily verified by solving the Eqs (9,10). Curiously, however, the plots of $g(x;s,1)$ for $s<s_{c}<0$ in Fig.9 formally \textit{look like} producing two distinct horizon radii [meaning $g(x;s,1)=0$ at two points on the positive $x-$axis] if we are ready to accept a contravention of the definition, $s>0$ or $q>0$.

To be sure, let us verify the behavior of the Hod function $H_{g}\left( w,\mu ,x,s\right) $ in the simplest case of coincident double horizons (Figs.8 and 9) at $x=x_{c}$ corresponding to the parameter value $s=s_{c}$. It yields, respectively%
\begin{equation}
H_{g}\left( 0,3,x_{c},s_{c}\right) =-64<0
\end{equation}%
\begin{equation}
H_{g}\left( 1,3,x_{c},s_{c}\right) =\frac{320}{137}>1,
\end{equation}%
which show that the inequality (32), or the hoop conjecture, is \textit{not} satisfied in either of the cases $w=0,1$ for $\mu =3$, hence these cases cannot be said to represent black holes.

Let us look at the curvature invariants. The Ricci scalar $R\equiv g^{\mu\nu}R_{\mu\nu}$ and the Kretschmann scalar $K\equiv R_{\mu \nu \alpha\beta }R^{\mu \nu \alpha \beta }$ for the metric $g(x;s,0)$ for $\mu =3$ turns out to be, respectively,
\begin{equation}
R=-\frac{16}{3q^{2}},K=\frac{211}{36q^{4}}.
\end{equation}%
These scalars for the function $g(x;s,1)$ for $\mu =3$ also show similar behavior
\begin{equation}
R=-\frac{16}{9q^{2}},K=\frac{14.4667}{q^{4}}
\end{equation}%
They are all finite, the Kretschmann scalar is positive but the negative Ricci scalars suggest that the new spacetime (36) has a topology akin to that of a wormhole\footnote{%
The most widely discussed Morris-Thorne wormhole \cite{Morris:1988} in standard coordinates has the metric, $d\tau ^{2}=-dt^{2}+\left( 1-\frac{m^{2}}{r^{2}}\right) ^{-1}dr^{2}+r^{2}(d\theta ^{2}+\sin ^{2}\theta d\psi ^{2})$ for which the Ricci scalar is $R=-\frac{m^{2}}{r^{4}}=-\frac{1}{d^{2}}$, where $d=r^{2}/m$ has the dimension of length.}.

Let us consider another case, $\mu =4,w=1$. We obtain the following values%
\begin{eqnarray}
x_{c} &=&\frac{7}{5},s_{c}=-\frac{3125}{62208}, \\
R &=&-\frac{35}{12q^{2}},K=\frac{12.7509}{q^{4}}.
\end{eqnarray}%
\begin{equation}
H_{g}(1,4,x_{c},s_{c})=-\frac{4375}{953}<0.
\end{equation}

It is possible to enumerate all the relevant quantitites for arbitrary values of $\mu$ and $w$ but clearly all the three cases ($\mu =3, w=0$), ($\mu =3, w=1$) and ($\mu =4, w=1$) above are sufficient to argue that the there are significant difficulties that speak against the proposed new class of solutions to be black holes, notably the occurrence of negative $s_{c}$, wormhole-type topology and of course, the violation of Thorne's hoop conjecture.

\begin{figure}[!ht]
  \centerline{\includegraphics[scale=1.05]{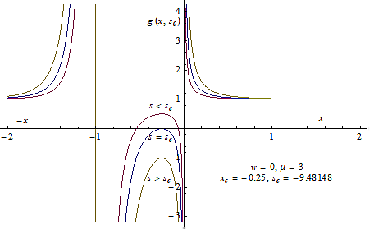}}
  \caption{Behavior of the new class of functions functions $g(x;s)$ for fixed values of the parameter $s<s_{c}$ and $w=0$, $\mu =3$. The plot shows no black hole for positive values of $x$.}
  \label{fig8}
\end{figure}

\begin{figure}[!ht]
  \centerline{\includegraphics[scale=1.05]{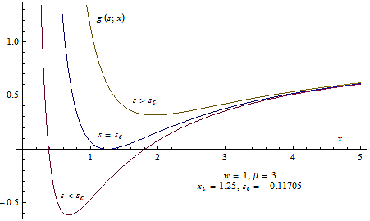}}
  \caption{Behavior of the new class of functions functions $g(x;s)$ for fixed values of the parameter $s<s_{c}$ and $w=1$, $\mu =3$. The plot looks like showing horizons for positive values of $x$ but is based on inadmissible $s<0.$}
  \label{fig9}
\end{figure}

\section{Equality of Hod and Misner-Sharp masses}
\label{sec:6}

To prove the equality of the two masses analytically would require the use of Einstein's field equations that equate the matter part $T_{\mu\nu}$ (on the right hand side) with the geometrical part $G_{\mu \nu}$ (on the left hand side). While we reserve this task for a separate publication, we currently exemplify the equality by using the Fan-Wang solutions including their proposed new class.

In an asymptotically flat spacetime with metric%
\begin{equation}
d\tau^{2} = -h_{tt}dt^{2} + h_{rr}dr^{2} + r^{2}(d\theta^{2} + \sin^{2}\theta d\psi^{2}),
\end{equation}%
the geometric Misner-Sharp quasilocal mass \cite{Misner:1973,Misner:1964,Hammad:2016} contained within a ball of radius $r=R$ is given by%
\begin{eqnarray}
m(r &\leq &R) = \frac{r}{2}(1-h^{\mu \nu }\partial _{\mu }r\partial _{\nu }r) \\
&=& \frac{r}{2}(1-h^{rr}\partial _{r}r\partial _{r}r)  \nonumber \\
&=& \left. \frac{r}{2}(1-h^{rr})\right\vert _{r=R}
\end{eqnarray}%
We want to show that this is equal to the Hod mass \ $M(r\leq R)$ in Eq.(19), to wit,%
$$
M(r\leq R) = M_{\infty} - E_{\scriptsize{\textmd{elec}}} = M_{\infty} - \frac{1}{8\pi}\int_{R}^{\infty }T_{0}^{0}4\pi r^{2}dr.
$$
Consider the generic metric (3) with the total ADM mass $M_{\infty }=M_{\scriptsize{\textmd{Sch}}}+M_{e}.$

\textbf{Case 1:} $M_{\scriptsize{\textmd{Sch}}}=0,M_{e}\neq 0$. Then identifying the metric components in (3) with those in (51), we have $M_{\infty }=M_{e}$ and%
\begin{equation}
h^{rr}=1-\frac{2M_{e}r^{\mu -1}}{\left( r^{\nu }+q^{\nu }\right) ^{\mu /\nu }},
\end{equation}%
\begin{eqnarray}
m(r\leq R)&=& \left. \frac{r}{2}(1-h^{rr})\right\vert _{r=R} = \left. \left(
\frac{r}{2}\right) \frac{2M_{e}r^{\mu -1}}{\left( r^{\nu }+q^{\nu }\right)
^{\mu /\nu }}\right\vert _{r=R} \nonumber \\
&&= M_{e}\left[ 1+\left( \frac{q}{R}\right)
^{\nu }\right] ^{-\frac{\mu }{\nu }}.
\end{eqnarray}%
This is precisely the expression $M(r\leq R)$ that can be found from Eq.(19)
putting the value of the integral from Eq.(20).

\textbf{Case 2:} $M_{\scriptsize{\textmd{Sch}}}\neq 0$, $M_{e}\neq 0$. Then identifying the metric components in (3) with those in (51), we have $M_{\infty }=M_{\scriptsize{\textmd{Sch}}}+M_{e}$ and

$$
h^{rr}=1-\frac{2M_{\scriptsize{\textmd{Sch}}}}{r}\frac{2M_{e}r^{\mu -1}}{\left( r^{\nu
}+q^{\nu }\right) ^{\mu /\nu }}. $$
\begin{eqnarray}
m(r \leq R)&=& \left. \frac{r}{2}(1-h^{rr})\right\vert _{r=R} \nonumber \\
&=& \left. \left( \frac{r}{2}\right) \left[ \frac{2M_{\scriptsize{\textmd{Sch}}}}{r}+\frac{2M_{e}r^{\mu -1}}{\left( r^{\nu }+q^{\nu }\right) ^{\mu /\nu }}\right] \right\vert _{r=R} \\
&=& M_{\scriptsize{\textmd{Sch}}}+M_{e}\left[ 1+\left( \frac{q}{R}\right) ^{\nu }\right] ^{-\frac{\mu }{\nu }}.
\end{eqnarray}%
This is precisely the expression $M(r\leq R)$ that can be found from Eq.(19) putting the value of the integral from Eq.(20).

\textbf{Case 3.} $M_{\scriptsize{\textmd{Sch}}}\neq 0$, $M_{e}\neq 0$. The new class of solution in Sec.5 yields the Misner-Sharp mass%
\begin{equation}
m(r\leq R)=\left. \frac{r}{2}(1-h^{rr})\right\vert _{r=R}=M_{\scriptsize{\textmd{Sch}}}+%
\frac{M_{e}R^{\mu }}{(R+q)^{\mu }}.
\end{equation}%
This is precisely the expression $M(r\leq R)$ that can be found from Eq.(19) putting the value of the integral from Eq.(37) together with $M_{\infty }=M_{\scriptsize{\textmd{Sch}}}+M_{e}$. Rescaling by $\frac{M_{e}}{R}=\frac{1}{2sx}$, $x=\frac{R}{q}$, we can obtain the respective Hod functions (23), (31) and (38).

\section{Conclusions}
\label{concl}

We believe that Thorne's famous hoop conjecture should be respected by static spherically symmetric black holes\footnote{%
Hod \cite{Hod:2020aa} has recently shown that the hoop conjecture valid for static black holes does not hold for spinning black holes and he proposed a new and unified "inverse hoop conjecture" \cite{Hod:2020ab} valid for spinning as well as static black holes. This new conjecture has been shown \cite{Nandi:2021a} to hold for black hole solutions belonging to a variety of extended gravitational theories including of course general relativity.}. Therefore, testing the validity of the conjecture for given spacetimes is testing whether they represent black holes or not. In our understanding, the result of the test can by no means be trivially predicted, particularly when the spacetimes contain \textit{arbitrary} exponents such as the ones considered above..

In this paper, the conjecture has been restated in terms of an extended Hod function $H_{f}$ corresponding to the generalized solutions, proposed by Fan and Wang \cite{Fan:2018}, with two arbitrary exponents $\mu $,$\nu $ coupled to nonlinear electrodynamics. These solutions generalize the seed black hole solutions of Bardeen \cite{Bardeen:1968} and Hayward \cite{Hayward:2006}. The function $H_{f}$ is based on Hod's suggestion \cite{Hod:2018} that the engulfed mass from all directions is the mass $M(r\leq R)$ enclosed within the hoop radius $R$, and not the asymptotic ADM mass $M_{\infty }$. The Hod function, first named and studied in \cite{Nandi:2020} for the seed solutions, encapulates the conjecture, which is then applied here to test the generalized solutions proposed in \cite{Fan:2018}.

In Sec.2, horizon conditions on metric functions $f$ in \cite{Fan:2018} across the critical values $x=x_{c}$, $s=s_{c}>0$ are stated in (13)-(15). These critical values are obtained by solving simultaneous Eqs.(9,10), which are reflected in Figs.(2-5) depicting the metric functions. It was argued after Eq.(16) [and also after Eq.(28)] how even a "no horizon" value of $s$ ($>s_{c}$) can mislead one to perceive a "horizon". This contradiction amounts to a violation of the conjecture if it is based on asymptotic mass as in Eq.(16). The Hod function removes such contradictions and resultant violations. The relevant functions $H_{f}$ for $M_{\scriptsize{\textmd{Sch}}}=0$ and $M_{\scriptsize{\textmd{Sch}}}\neq 0$ are worked out in Secs.3,4 respectively and the condition $0<H_{f}(\mu ,\nu ;x_{c},s)\leq 1$ for $s\leq s_{c}$ imposed by the conjecture on the function $H_{f}$, is graphically shown to hold for all values of $\mu$, $\nu$ (some are shown in Figs.6,7). Thus the corresponding generalized solutions are truly black holes like their seed solutions in \cite{Bardeen:1968,Hayward:2006}. Clearly, $H_{f}$ provides a beautiful tool for testing the validity of the generalized solutions as black holes.

The \textit{new} class of solutions proposed by Fan and Wang (Sec.5), which are not generalizations of seed black hole solutions, however show intriguingly peculiar properties. It technically exhibits horizons (Fig.9) although the plots are based on $s_{c}<0$, which is \textit{inconsistent} with the stipulations, $\alpha >0,q>0$ [see Eq.(40)]. The curvatures are finite everywhere but the Ricci scalar turns out to be \textit{negative} meaning that the spacetime has a wormhole-like topology and finally, the hoop conjecture is \textit{violated}. All these factors speak against the new class of solutions to be called as black holes. However, if at least the inconsistency with the requirement $s>0$ could somehow be resolved, the class of new solutions (36) could pass for new "hybrid black holes" with a wormhole topology and violation of hoop conjecture then would not appear as an obstacle because of probable energy condition violations in the spacetime.

We also exemplified (Sec.6) that the Hod mass $M(r\leq R)$ is exactly the matter counterpart of the Misner-Sharp \cite{Misner:1973,Misner:1964} geometrical quasilocal mass of general relativity. Thus, we emphasize that the adjudication of black holes (or otherwise) based on $H_{f}$ has the authority of general relativity.

We believe that the analyses presented above should be useful from the methodological point of view as well since the relevant Hod functions can be developed to study the conjecture for various other complicated situations such as the spacetimes coupled to scalar field or dilatonic field or in modified gravity theories. Work is in progress.

\section*{Acknowledgments}

We thank an anonymous reviewer for his/her constructive comments.

\end{document}